\documentclass[preprint,showpacs,preprintnumbers,amsmath,amssymb,nofootinbib]{revtex4}

\usepackage{epsf}

\newcommand{\be}{\begin{equation}}
\newcommand{\ee}{\end{equation}}
\newcommand{\ba}{\begin{eqnarray}}
\newcommand{\ea}{\end{eqnarray}}

\begin{document}

\title{Observational signatures of the weak lensing magnification of supernovae}
\author{Yun Wang}
\affiliation{Department of Physics \& Astronomy, University of Oklahoma, 
Norman, OK 73019 USA.\\
wang@nhn.ou.edu}

\date{\today}
{\begin{abstract} 
Due to the deflection of light by density fluctuations along the
line of sight, weak lensing is an unavoidable systematic uncertainty
in the use of type Ia supernovae (SNe Ia) as cosmological
distance indicators. 
We derive the expected weak lensing
signatures of SNe Ia by convolving the intrinsic 
distribution in SN Ia peak luminosity with magnification
distributions of point sources. 
We analyze current SN Ia data, and find 
marginal evidence for weak lensing effects. 
The statistics is poor 
because of the small number of observed SNe Ia. 
Future observational data will allow unambiguous detection 
of the weak lensing effect of SNe Ia.
The observational signatures of weak lensing of SNe Ia
that we have derived provide useful templates
with which future data can be compared.
\end{abstract}}




\maketitle
 
\section{Introduction}

The use of type Ia supernovae (SNe Ia) as cosmological
distance indicators has become fundamental in observational 
cosmology \citep{Garna98,Riess98,Perl99,Knop03,Tonry03,Riess04}.
Although SNe Ia can be calibrated to be good standard 
candles \citep{Phillips93,Riess95},
they can be affected by systematic uncertainties. 
These include possible evolution in the intrinsic SN Ia peak brightness 
with time \citep{Drell00}, 
weak lensing of SNe Ia
\citep{Kantow95,frieman97,Wamb97,Holz98,ms99,Wang99,Valageas00,MJ00,Barber00,Premadi01}, 
and possible extinction by gray dust \citep{Aguirre99}.

Weak lensing effect is an unavoidable systematic uncertainty
of SNe Ia as cosmological standard candles, simply because there are fluctuations
in the matter distribution in our universe, and they deflect the
light from SNe Ia (causing either demagnification or magnification). 

In Sec.2, we derive the expected weak lensing
signatures of SNe Ia by convolving the intrinsic 
distribution in SN Ia peak luminosity with magnification
distributions of point sources.
In Sec.3, we use current SN Ia data to show that weak lensing effect may 
have already begun to set in.
Sec.4 contains a brief summary and discussions.

\section{Signatures of weak lensing}

The observed flux from a SN Ia is 
\be
f=\mu L_{int},
\ee
where $L_{int}$ is the intrinsic brightness of the SN Ia,
and $\mu$ is the magnification due to intervening matter.
Note that $\mu$ and $L_{int}$ are statistically independent.
The probability density distribution (pdf) of the product of two
statistically independent variables can be found
given the pdf of each variable (for example,
see \cite{Lupton}).

We find that the pdf of the observed flux $f$ is given by
\be
\label{eq:convolve}
p(f) = \int_0^{L_{int}^{max}} \frac{dL_{int}}{L_{int}}
\,g\left(L_{int}\right)\,
p\left(\frac{f}{L_{int}}\right) ,
\ee
where $g(L_{int})$ is the pdf of the intrinsic
peak brightness of SNe Ia, $p(\mu)$ is the pdf of the
magnification of SNe Ia.
The upper limit of the integration,
$L_{int}^{max}=f/\mu_{min}$, results from
$\mu= f/L_{int} \geq \mu_{min}$.

A definitive measurement of $g(L_{int})$ will
require a much greater number of well measured SNe Ia
at low $z$ than is available at present.
Since $g(L_{int})$
is not sufficiently well determined at present, we present
our results for two different $g(L_{int})$'s:
Gaussian in flux and Gaussian in magnitude.

The $p(\mu)$'s can be computed numerically using
cosmological volume N-body simulations
(see for example, \cite{Wamb97,Barber00,Premadi01,Vale03}).
We can derive the $p(\mu)$ for an arbitrary cosmological
model by using the universal probability distribution 
function (UPDF) of weak lensing amplification \cite{Wang99,WangHolzMunshi},
with the corrected definition of the minimum convergence
\citep{Wangalphaeta},
\be
\label{eq:kapdef}
\tilde{\kappa}_{min}(z) 
= - \frac{3}{2} \, \frac{\Omega_m (1+z)}{cH_0^{-1}}
\int_0^z dz'\, \frac{(1+z')^2}{E(z')}\, \frac{r(z')}{r(z)} 
\,[\lambda(z)-\lambda(z')],
\ee
where $r(z)$ is the comoving distance in a smooth universe,
\[
E(z) \equiv \sqrt{\Omega_m (1+z)^3 +\Omega_X \rho_X(z)/\rho_X(0) 
+ \Omega_k (1+z)^2},
\] 
with $\rho_X(z)$ denoting the dark energy
density. 
The affine parameter 
\be
\lambda(z) = cH_0^{-1}\, \int_0^z \frac{dz'}{(1+z')^2\, E(z')}.
\ee
Note that $\mu_{min}= 1/(1-\tilde{\kappa}_{min})^2$.
We have used a modified UPDF \citep{Wang04}, with the corrected 
minimum convergence and extended to high magnifications. The numerical
simulation data of $p(\mu)$ is converted to the modified UPDF,
pdf of the reduced convergence $\eta$, 
\be
P(\eta)=\frac{1}{1+\eta^2} \, \exp\left[ - \left(
\frac{\eta-\eta_{peak}}{w \eta^q}  \right)^2 \right],
\ee
where $\eta= 1+ (\mu-1)/|\mu_{min}-1|$. The parameters of the UPDF,
$\eta_{peak}$, $w$, and $q$ are only functions of the variance of
$\eta$,  $\xi_{\eta}$,
which absorbs all the cosmological model dependence.
The functions $\eta_{peak}(\xi_{\eta})$, $w(\xi_{\eta})$, and $q(\xi_{\eta})$ 
are extracted from the numerical simulation data.
For an arbitrary cosmological model, one can readily compute 
$\xi_{\eta}$ \cite{WangHolzMunshi},
and then the UPDF (and hence $p(\mu)$) can be computed.

Fig.1 shows the prediction of the observed flux distributions of SNe Ia, 
with magnification distribution $p(\mu)$ given by a $\Lambda$CDM model 
($\Omega_m$=0.3, $\Omega_{\Lambda}=0.7$) at $z=1.4$ (top panel)
and $z=2$ (bottom panel) respectively.
We have assumed that the intrinsic peak brightness
distribution, $g(L_{int})$, is Gaussian with a rms variance of
0.193 (in units of the mean flux, and chosen to be the same
as the $0.02 \leq z \leq 0.1$ subset of the Riess 2004 sample).

\begin{figure}[h]
\vskip-0.3cm \centerline{\epsffile{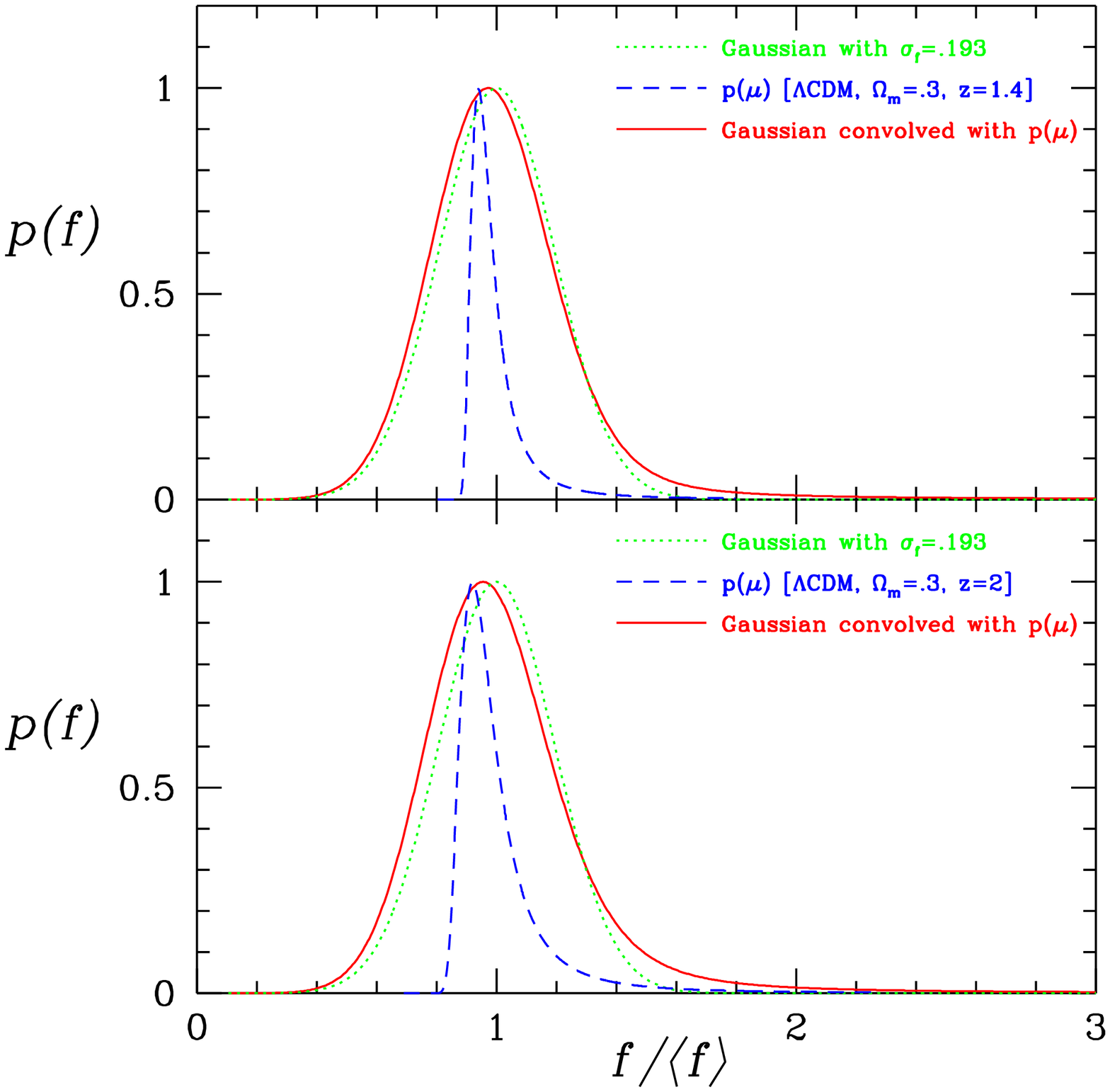}}
\caption[1]{\footnotesize%
Prediction of the observed flux distributions 
of SNe Ia for magnification distribution $p(\mu)$ given by a $\Lambda$CDM model 
($\Omega_m$=0.3, $\Omega_{\Lambda}=0.7$) at $z=1.4$ (top panel)
and $z=2$ (bottom panel) respectively.
We have assumed that the intrinsic peak brightness
distribution, $g(L_{int})$, is Gaussian with a rms variance of
0.193 (in units of the mean flux). }
\end{figure}

Fig.2 is the same as Fig.1, except here we
have assumed that $g(L_{int})$ is Gaussian
in magnitude, with a rms variance of 0.213 mag
(chosen to be the same as the $0.02 \leq z \leq 0.1$ 
subset of the Riess 2004 sample).

\begin{figure}[h]
\vskip-0.3cm \centerline{\epsffile{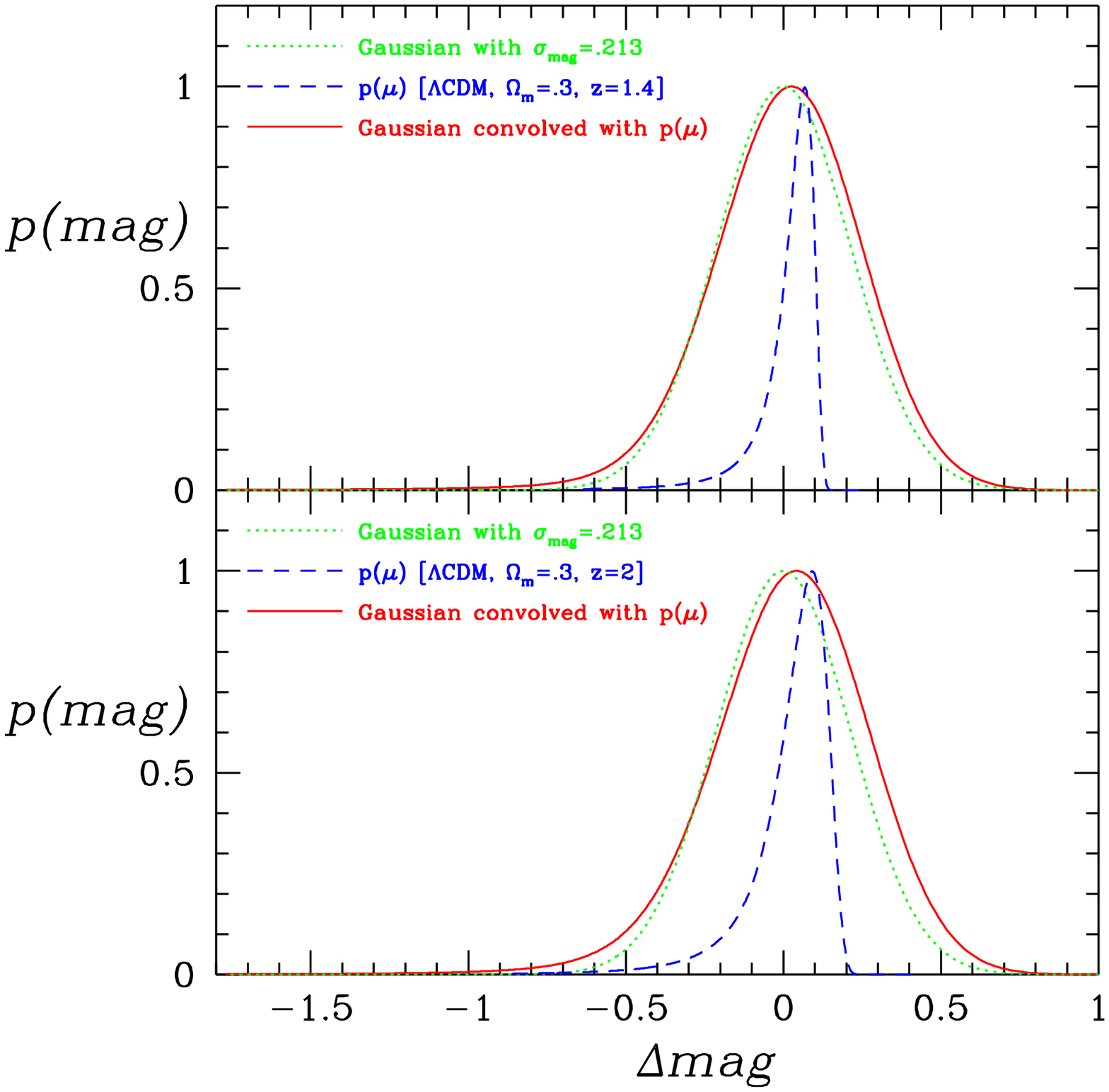}}
\caption[1]{\footnotesize%
Same as Fig.1, except here we
have assumed that the intrinsic peak brightness
distribution, $g(L_{int})$, is Gaussian
in magnitude, with a rms variance of 0.213 mag. }
\end{figure}

Clearly, there are two signatures of the weak lensing of SNe Ia
in the observed brightness distribution of SNe Ia.
The first signature is the presence of a non-Gaussian tail at 
the bright end, which is due to the high magnification tail of
the magnification distribution.
The second signature is the slight shift of the
peak toward the faint end (compared to the pdf
of the intrinsic SN Ia peak brightness), which is
due to $p(\mu)$ peaking at $\mu <1$
(demagnification) because the universe is mostly empty.
As the redshift of the observed SNe Ia increases,
the non-Gaussian tail at the bright end will grow larger,
while the peak will shift further toward the faint end (see
Figs.1-2).

If the distribution of the intrinsic SN Ia peak brightness
is Gaussian in flux, the dominant signature of weak lensing
is the presence of the high magnification tail in flux.
If the distribution of the intrinsic SN Ia peak brightness
is Gaussian in magnitude, the dominant signature of weak lensing
is the shift of the peak of observed magnitude distribution
toward the faint end due to demagnification. This is as expected, since
the magnitude scale stretches out the distribution at
small flux, and compresses the distribution at large flux.

\section{Evidence of weak lensing in current supernova data}

We use the Riess sample \citep{Riess04} to explore
possible weak lensing in current SN Ia data,
as this sample contains the largest number of
SNe Ia at $z>1$ that are publicly available.

Our high $z$ subset consists of 
63 SNe Ia from the Riess sample with with $0.5 \leq z \leq 1.4$.
Our low $z$ subset consists of 
47 SNe Ia from the Riess sample with $0.02 \leq z \leq 0.1$. 
Table 1 shows the redshift distribution of the 63
SNe Ia in the high $z$ subset.
To enable meaningful comparison of the high $z$ and
low $z$ samples, the bestfit cosmological
model has been subtracted from the brightnesses of all the
SNe Ia in both samples. The bestfit cosmological
model (found by allowing the dark energy density to be a free function
given by 4 parameters) has been obtained via flux-averaging of the gold set of
157 SNe Ia of the Riess sample, combined with CMB and
galaxy survey data \cite{WangTegmark}, hence it should have very
weak dependence on the mean brightnesses of the low $z$ 
and high $z$ SN Ia samples.

\begin{center}
Table 1\\
{\footnotesize{Redshift distribution of SNe Ia in the high $z$ subset}}

{\footnotesize
\begin{tabular}{|cccc|}
\tableline              
$z$ & [.5, .7] & [.7, .9] & [.9, 1.4]\\  
\tableline 
number & 27    & 21       & 15 \\   
\tableline
\end{tabular}
}
\end{center}

Fig.3 shows the flux distributions of 63 SNe Ia
with $0.5 \leq z \leq 1.4$ (top panel) and
47 SNe Ia with $0.02 \leq z \leq 0.1$ (bottom
panel). The bestfit cosmological model obtained by \cite{WangTegmark}
using flux-averaging (which minimizes weak lensing
effect, see \cite{Wang00b,WangMukherjee}) 
has been subtracted.

\begin{figure}[h]
\vskip-0.3cm \centerline{\epsffile{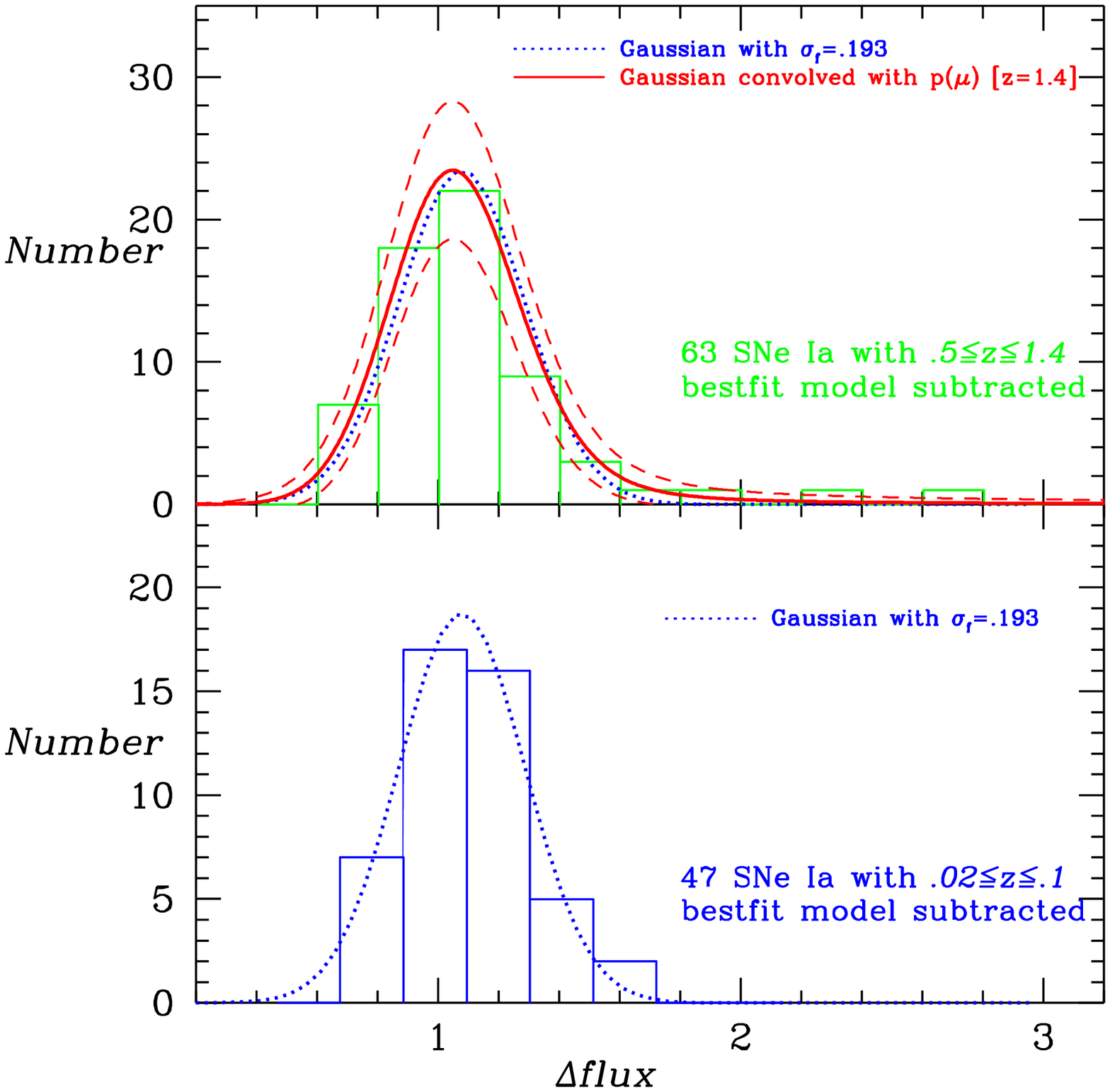}}
\caption[1]{\footnotesize%
The flux distributions of 63 SNe Ia
with $0.5 \leq z \leq 1.4$ (top panel) and
47 SNe Ia with $0.02 \leq z \leq 0.1$ (bottom
panel). The bestfit model obtained by \cite{WangTegmark}
using flux-averaging (which minimizes weak lensing
effect) has been subtracted.
Clearly, the distribution of the low $z$ SNe Ia is
consistent with Gaussian, while the high $z$
SNe Ia seem to show both signatures of weak lensing
(high magnification tail and demagnification shift of the 
peak to smaller flux). The error bars indicate the Poisson
noise of the weak lensing prediction (solid curve). }
\end{figure}

Fig.4 is the same as Fig.3, except here we have fitted
the distribution of the low $z$ SNe Ia to a Gaussian 
in magnitude, and have binned the high $z$ SNe Ia
in magnitude as well for comparison.

\begin{figure}[h]
\vskip-0.3cm \centerline{\epsffile{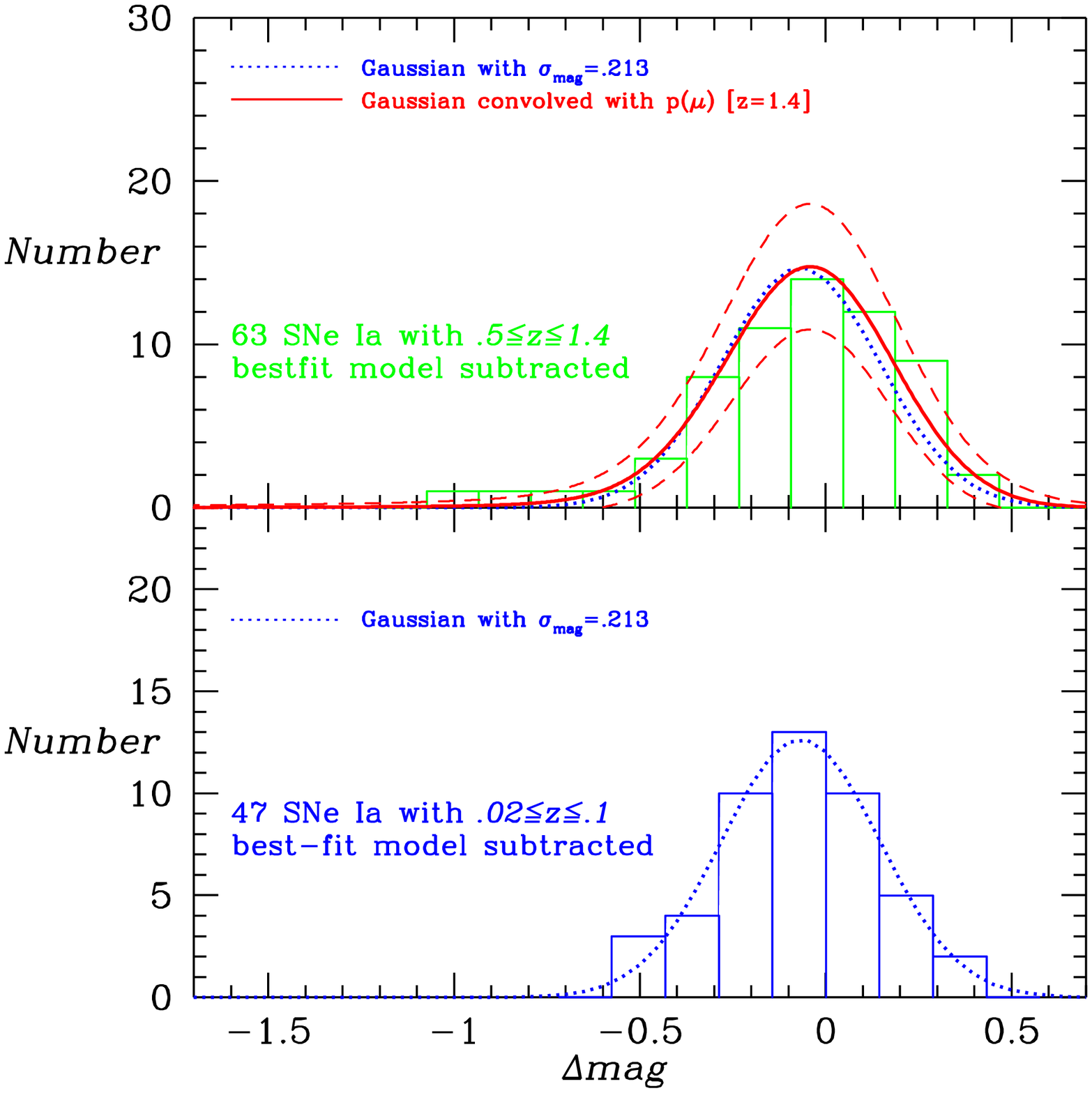}}
\caption[1]{\footnotesize%
Same as Fig.3, except here we have fitted
the distribution of the low $z$ SNe Ia to a Gaussian 
in magnitude, and have binned the high $z$ SNe Ia
in magnitude as well for comparison.
}
\end{figure}

The top panels of Figs.3 and 4 show the predicted 
distributions of SN Ia peak brightness, 
obtained by convolving the bestfit
Gaussian distribution at low $z$ (dotted line)
with $p(\mu)$ for the bestfit cosmological model with  
$z=1.4$ (solid line), and the Poisson noise
expected from the finite number of SNe 
Ia in each bin (dashed lines).

Clearly, the distribution of the low $z$ SNe Ia is
consistent with Gaussian in both flux and magnitude, 
while the high $z$
SNe Ia seem to show both signatures of weak lensing:
the high magnification tail at bright end
and the demagnification shift of the peak toward
the faint end.
Within the uncertainties of the Poisson noise,
the flux and magnitude distributions of the high $z$
SNe Ia are roughly consistent with the upper bound set
by the maximal expected amount of weak lensing 
in the best fit model.

Both the presence of the bright end tail
and the shift of the peak toward the faint end are
more pronounced than the predictions based on standard
lensing magnifications (see Figs.3-4). However, since the number of SNe Ia
at high $z$ is still small, the difference may
have resulted from statistical fluctuations
(Poisson noise).

Table 2 compares the high $z$ and low $z$ SNe Ia plotted in
Figs.3 and 4. The mean brightness and skewness of
the distributions have been calculated for both 
flux and magnitude distributions.  
The rms variance of the skewness $S_3$
for a Gaussian distribution is 
$\sigma_{S_3}^G \simeq \sqrt{6/N}$ \citep{Press}, 
where $N$ is the number of SNe Ia in the subset.

\begin{center}
Table 2\\
{\footnotesize{Comparison of high $z$ and low $z$ SNe Ia}}

{\footnotesize
\begin{tabular}{|lcccc|}
\tableline              
$z$ of SNe  & $\langle f \rangle^a$  &  $S_3 \pm \sigma_{S_3}^G$(flux)  
& $\langle mag \rangle^b$ & $S_3 \pm \sigma_{S_3}^G$(mag)\\  
\tableline 
$.5\leq z\leq 1.4 $     &   1.115  &  $2.12 \pm 0.31$  
& $-.0763$  & $-1.06 \pm 0.31$ \\  
$.5\leq z\leq 1.4 ^c$  &   1.055   &  $0.66 \pm 0.32$   
& $-.0352$  & $-0.25 \pm  0.32$\\ 
$.02\leq z\leq .1$     &   1.096  &    $0.60 \pm  0.36$   
 & $-.0805$  & $-0.19 \pm 0.36$  \\     
\tableline
\end{tabular}
}
\end{center}
{\footnotesize{$^a$ in units of the flux in the bestfit cosmological
model obtained via flux-averaging in \cite{WangTegmark}.\\
$^b$ defined to be $-2.5 \log(flux)$, where $flux$ is 
the dimensionless flux defined above.\\
$^c$ excluding the three brightest SNe listed in Table 3.}}

The high $z$ and low $z$ SNe Ia differ little
in mean brightness; this is as expected since the bestfit
cosmological model obtained using the gold set
of 157 SNe Ia of the Riess sample (together with CMB and galaxy clustering
data) has been subtracted.\footnote{Flux-averaging 
has been used in obtaining the bestfit model in order to minimize
the bias due to weak lensing \cite{WangTegmark}.} 
In the absence of weak lensing, one would expect that
the low $z$ and high $z$ SN Ia samples have similar distributions,
hence similar values of the skewness $S_3$.
If the intrinsic distribution of SN Ia brightnesses were Gaussian
in flux (magnitude), one would expect $S_3=0\pm \sigma_{S_3}^G$ for 
the data in flux (magnitude). 
Table 2 shows that:
(1) The high $z$ SN Ia sample has a significantly
larger skewness $S_3$ (for both flux and
magnitude distributions) than the low $z$ sample; 
(2) The large skewness
of the high $z$ sample is primarily due to the three brightest SNe Ia.
Both these points are consistent with
the signatures of weak lensing discussed in Sec.2.

Table 3 lists the three brightest SNe Ia in the
bright end tails of
Figs.3 and 4. All three SNe Ia are in the ``gold''
sample of \cite{Riess04}. The last column in Table 3
lists the possible range of magnification $\mu$
for each SN Ia, [$(f - df)/\langle f\rangle$, $(f + df)/\langle f\rangle$],
with $df/f = \sinh \left(\sigma_{\mu_0}\ln 10/2.5\right)$, 
and $\langle f\rangle$ given by the low $z$ 
subset.
Note that the possible magnification of SN1998I 
has a very large uncertainty, because its distance
modulus has a very large uncertainty
in the Riess sample: $\mu_0= 42.91\pm 0.81$,
which correspond to a flux uncertainty of about 80\%.

\begin{center}
Table 3\\
{\footnotesize{Three brightest SNe Ia}}

{\footnotesize
\begin{tabular}{|lcccc|}
\tableline 
SN  & $z$   & $\mu_0$   & $\sigma_{\mu_0}$ &  possible $\mu$\\  
\tableline 
SN1997as  &   0.508  &   41.64   &  0.35  &    [1.42, 2.78] \\        
SN2000eg  &   0.540  &   41.96   &  0.41  &    [1.10, 2.50]\\     
SN1998I   &   0.886  &   42.91   &  0.81  &    [0.44, 4.40] \\      
\tableline
\end{tabular}
}
\end{center}

Note that the bestfit cosmological model \cite{WangTegmark}
was obtained using the gold set of 157 SNe Ia of the Riess sample \cite{Riess04}, 
flux-averaged and combined with CMB and galaxy clustering data.
The mean brightnesses of the low $z$ and high $z$ samples considered
in this paper were {\it not} used in deriving the bestfit cosmological
model (which is given by 6 parameters).
The fact that the low $z$ and high $z$ samples have about the same
mean brightnesses shows the validity of the bestfit cosmological model,
since weak lensing does {\it not} change the mean brightness of teh high
$z$ sample due to flux conservation.

The second row in Table 2 shows that
if we exclude the three brightest SNe listed in Table 3,
the skewness of the high $z$ sample drops to about the
same as that of the low $z$ sample.
However, the mean brightness of the high
$z$ sample drops below that of the low $z$
sample, such that the high $z$ sample is about 4\% 
fainter in flux than the low $z$ sample.
This suggests that the three brightest SNe in the high z sample
are probably not outlyers, but may belong to the high magnification
tail of $p(\mu)$.\footnote{Note that weak lensing leads to a continuous
probability distribution function (pdf) in magnification  
with a tail at high magnifications which are often associated 
with strong lensing.
However, strong lensing usually refers to lensing that
results in multiple images of a point source \citep{Turner84}. 
The high magnification tail of a weak lensing pdf corresponds to
the bending of light of a point source due to galaxies that 
results in a {\it single} magnified image \citep{Wamb98}, hence it is
technically still weak lensing magnification, and not strong lensing.}

Finally, we do a Kolmogorov-Smirnov test to assess whether the low $z$
and high $z$ SNe Ia could be from the same brightness distribution
(i.e., no weak lensing of the high $z$ sample).
Table 4 shows the Kolmogorov-Smirnov test of the low $z$ versus high $z$
SN Ia sample, and the high $z$ SN Ia sample compared to a Gaussian
in magnitude (with $\sigma_{mag}=0.213$) convolved with $p(\mu)$ at
$z=1.4$ (as plotted in Fig.4a). We have chosen the later distribution
for comparison with the high $z$ sample, because the low $z$ sample
appears slightly more Gaussian in magnitude (with smaller $S_3$,
see Table 2).

\begin{center}
Table 4\\
{\footnotesize{Kolmogorov-Smirnov test}}

{\footnotesize
\begin{tabular}{|l|cc|cc|}
\tableline 
subtracted cosmological model	&   bestfit model &        & 
($\Omega_m=0.27$, $\Omega_\Lambda=0.73$)& \\
\hline
  			&    D            &  prob. &    D            &  prob. \\
low $z$ and high $z$ sample & 0.153  & 0.522  & 0.184 &    0.291 \\
high $z$ sample and $p(mag)^a$ & 0.096 &  0.585 & 0.104 &  0.482 \\
\tableline
\end{tabular}
}
\end{center}
{\footnotesize{$^a$ a Gaussian
in magnitude (with $\sigma_{mag}=0.213$) convolved with $p(\mu)$ at
$z=1.4$, see Eq.(\ref{eq:convolve}).}}

The Kolmogorov-Smirnov test gives $D$, the maximum value of the
absolute difference between two cumulative distribution functions, 
and $prob$, the probability that $D > observed$. 
Small values of $prob$ show that the two
distributions are significantly different.
Table 4 shows that current data do not yield results of high statistical significance,
however, the high $z$ SN Ia sample is more consistent with
a lensed distribution than with the low $z$ SN Ia sample.
Changing the subtracted cosmological model from the bestfit model
to a popular model with $\Omega_m=0.27$, $\Omega_\Lambda=0.73$
does not change our results qualitatively.

\section{Summary and Discussion}

We have derived the expected weak lensing signatures of SNe Ia
in the distribution of observed SN Ia peak brightness,
the presence of a high magnification tail
at the bright end of the distribution, and
the demagnification shift of the peak of the
distribution toward the faint end (see Figs.1-2).

Our method is complementary to that of
\cite{Metcalf01,Williams}; they
use the correlation between foreground galaxies and 
supernova brightnesses to detect weak lensing,
while our method only uses the statistics of supernova 
brightnesses and does not depend on the observation 
of foreground galaxies.

We have compared 63 high $z$ SNe Ia 
($0.5 \leq z \leq 1.4$) with 47 low $z$ SNe Ia
($0.02 \leq z \leq 0.1$) from the Riess sample
\citep{Riess04}. We find that 
the observed flux and magnitude
distributions of the high $z$ sample
are roughly consisitent with the maximal expected amount
of weak lensing magnification (see Figs.3-4), within
the Poisson noise due to the small number
of SNe Ia in each bin.

We have identified the three brightest SNe Ia 
in the high $z$ subset ($0.5 \leq z \leq 1.4$)
of the Riess sample, and estimated a
possible range of magnification for each SN Ia (Table 3).
Observational follow-up of the regions near these
SNe Ia may show whether these SNe Ia have indeed
been magnified. 
Note that selection effects should not be important here, since the
three brightest SNe are at intermediate redshifts (where the fainter SNe 
are not close to the detection limits of the surveys).

Our results are consistent with those of
\cite{Williams,Bassett04}. \cite{Williams} found that brighter SNe Ia
are preferentially found behind regions which are
overdense in foreground galaxies, as expected
in weak lensing.
\cite{Bassett04} found tentative
evidence for a deviation from the reciprocity relation
between the angular diameter distance and the luminosity
distance, which could be due to the
brightening of SNe Ia due to lensing.

Although the observed flux/magnitude
distribution of the high $z$ SN Ia sample
deviates from a Gaussian in ways that
are qualitatively consisitent with the weak lensing
effects (allowing for Poisson noise), it is possible
that these deviations could be due to other unidentified
systematic effects. However, it is important to note that
if we remove the three brightest SNe Ia in the
high $z$ sample, the distributions become more
Gaussian, but the mean becomes biased (see Table 2).
Therefore, these three SNe Ia might not be outliers,
and should be included in the data analysis.

Fortunately, the non-Gaussianity of the high $z$ SN Ia flux/magnitude
distribution (regardless of its origin -- weak lensing or
some other systematic effect) does {\it not} seem
to alter the mean of the distribution (compared to
the low $z$ sample).
This means that we can, and should, use
flux averaging in order to obtain
unbiased estimates of cosmological parameters
\citep{Wang00b,WangMukherjee}.

Our results explain the difference between
the cosmological constraints found by \cite{Riess04}
and \cite{WangTegmark} for the same model assumptions 
(see Fig.10 of \cite{Riess04} and Fig.2(a) of 
\cite{WangTegmark}).
\cite{WangTegmark} used flux-averaging in their
likelihood analysis; \cite{Riess04} did not.

As more SNe Ia are discovered at high $z$,
it becomes increasingly important to minimize the
effect of weak lensing (or other non-Gaussian systematic 
effects that conserve flux) by flux-averaging 
\citep{Wang00b,WangMukherjee}\footnote{If the 
distribution of the intrinsic peak brightness
of SNe Ia is Gaussian in magnitude, then
flux-averaging would introduce a small bias
of $-\sigma_{mag}^2 \ln 10/5 $\citep{Wang00a},
which needs to taken into account in the data analysis.}
in using SNe Ia to probe cosmology. 

It seems that weak lensing effects, or some other systematic
effect that mimics weak lensing qualitatively,
may have begun to set in (see Figs.3-4 and Tables 2 \& 4).
However, the statistics is poor because of the small
number of observed SNe Ia. 
Future observational data from current and planned 
SN Ia surveys will allow unambiguous detection
of the weak lensing effect of SNe Ia.
The observational signatures of weak lensing of SNe Ia
that we have derived (see Figs.1-2) provide useful templates
with which future data can be compared.


{\bf Public software:} A Fortran code that uses flux-averaging statistics 
to compute the likelihood of an arbitrary 
dark energy model (given the SN Ia data from \cite{Riess04})  
can be found at $http://www.nhn.ou.edu/\sim wang/SNcode/$.

\acknowledgements

This work was supported in part by 
NSF CAREER grant AST-0094335.
I thank David Branch, Alex Filippenko, Daniel Holz, 
and Max Tegmark for helpful discussions.

\end{document}